\documentstyle[11pt,newpasp,twoside,epsf]{article} 
\def\edcomment#1{\iffalse\marginpar{\raggedright\sl#1\/}\else\relax\fi} 
\reversemarginpar

\begin{document}

\title{Rotation and abundance anomalies in blue horizontal-branch stars}
\author{Bradford Behr} 
\affil{McDonald Observatory, University of Texas at Austin, 1 University Station C1400,
	Austin TX 78712-0259, bbb@astro.as.utexas.edu}

\begin{abstract} 
To address the puzzling photometric properties of horizontal branch stars in Galactic GCs, several
different groups have undertaken detailed spectroscopic analyses of individual blue HB stars. Hotter
BHB stars show strong metal enhancement and helium depletion, likely due to atomic diffusion, and
slow rotation velocities, in contrast to the cooler BHB stars, which have a bimodal distribution of
rotation speeds --- some of them much faster than expected --- but no anomalous abundances. I review
the observational results to date, and discuss possible explanations and ramifications of these
abundance and rotation characteristics.
\end{abstract} 

\section{Introduction}

Despite extensive observations and detailed theoretical modelling of post-main-sequence stellar
evolution, we still cannot fully explain the color distributions and detailed properties of horizontal
branch (HB) stars in globular clusters. The relative number of red versus blue HB stars is primarily a
function of a cluster's metallicity, but the wide variety of HB color morphologies found among Galactic
GCs with similar metallicities strongly suggests that some other factor, or combination of factors, must
also be at work. Cluster age, helium abundance, stellar kinematics, and binarity have all been
considered, but a clearly defined ``second parameter'' has not yet been identified. Furthermore,
high-precision photometry and spectrophotometry reveals further anomalies in HB populations: narrow
``gaps'' in the distribution of stars along the HB locus (Ferraro et al. 1998), and temperature
ranges over which stars are consistently ``overluminous'' (Grundahl et al. 1999), or have
unexpectedly low derived surface gravities (Moehler 2001), compared to canonical HB loci.
Evidently, our models do not fully account for all of the relevant properties of these stars.

In order to better understand the characteristics of these stars, several research groups have
undertaken detailed high-resolution spectroscopic observations of individual HB stars in metal-poor
globular clusters and the field halo population. These investigations have discovered significant
variation in both the photospheric chemical abundances and the rotation velocities of these stars,
correlating with position along the HB. In this paper, I review the observational work that has been
reported to date, discuss possible explanations for these rotation and abundance anomalies, and explore
their potential implications for stellar evolution and cluster photometry and kinematics.

\section{Abundance and rotation observations}

\begin{figure}
\plotone{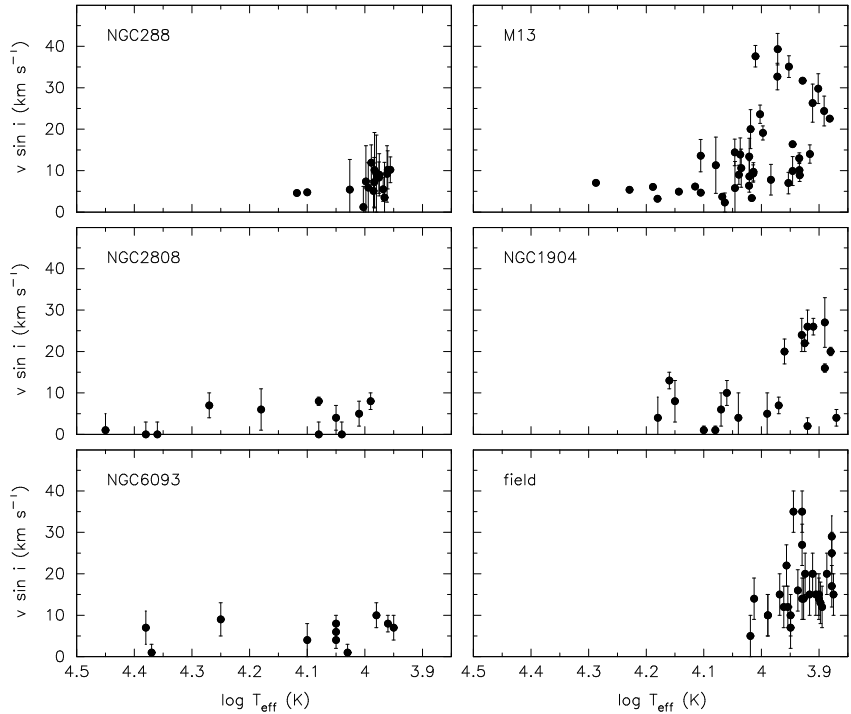}
\caption{Projected rotation velocities for BHB stars as a function of HB position. Bimodal distributions
in $v \sin i$ are seen in M13, NGC~1904, and the field, but not in the other clusters. Data
are from Peterson et al. (1995), Behr (2000c), Kinman et al. (2000), and Recio-Blanco et
al. (2002).}
\end{figure}

The pioneering high-resolution spectroscopic studies of HB stars were done by Peterson and
collaborators (1983ab, 1985ab, 1995), who observed cooler BHB stars in M3, M4, M5, M13,
NGC~288, and the nearby field population. They determined projected rotation velocities ($v \sin i$)
from the broadening of the metal absorption lines in their spectra, and found several of their target
stars spinning as fast as $40 \, {\rm km} \, {\rm s}^{-1}$. Such fast rotation was unexpected,
given that these HB stars had evolved from G dwarfs, which lose most of their initial angular
momentum to magnetically-coupled winds over their main-sequence lifetimes. Only a fraction of the
stars appeared to be spinning so fast; assuming random orientation of rotation axes, M13 and the
metal-poor field BHB population appeared to have a bimodal underlying distribution of actual rotation
velocities $v_{\rm rot}$, with roughly one-third of the stars at $v_{\rm rot} \simeq 40 \, {\rm km}
\, {\rm s}^{-1}$ and the other two-thirds at $v_{\rm rot} \simeq 15$--$20 \, {\rm km} \, {\rm
s}^{-1}$. The other clusters seemed to possess only the slowly-rotating stars. Cohen \& McCarthy
(1997) found similar $v_{\rm rot}$ bimodality among the cool BHB stars of M92, and Kinman et al.
(2000) enlarged the sample of field BHB stars, confirming the existence of both fast and slow rotators
among this population. Extending the observations in M13 to higher $T_{\rm eff}$, Behr et al. (2000a)
found that the anomalous fast rotators do not appear above $T_{\rm eff} \simeq 11500 \, {\rm K}$,
as all of the hotter BHB stars exhibit very slow rotation, $v \sin i < 8 \, {\rm km} \, {\rm s}^{-1}$.
A similar temperature dependence was subsequently found in M15 (Behr et al. 2000b) and NGC~1904
(Recio-Blanco et al. 2002). Recio-Blanco also found additional clusters, NGC~2808 and NGC~6093,
which appear to possess only the slowly-rotating BHB stars. Figure 1 shows a representative sample
of the BHB rotation measurements reported to date, illustrating the variety of $v \sin i$ distributions.

\begin{figure}
\plottwo{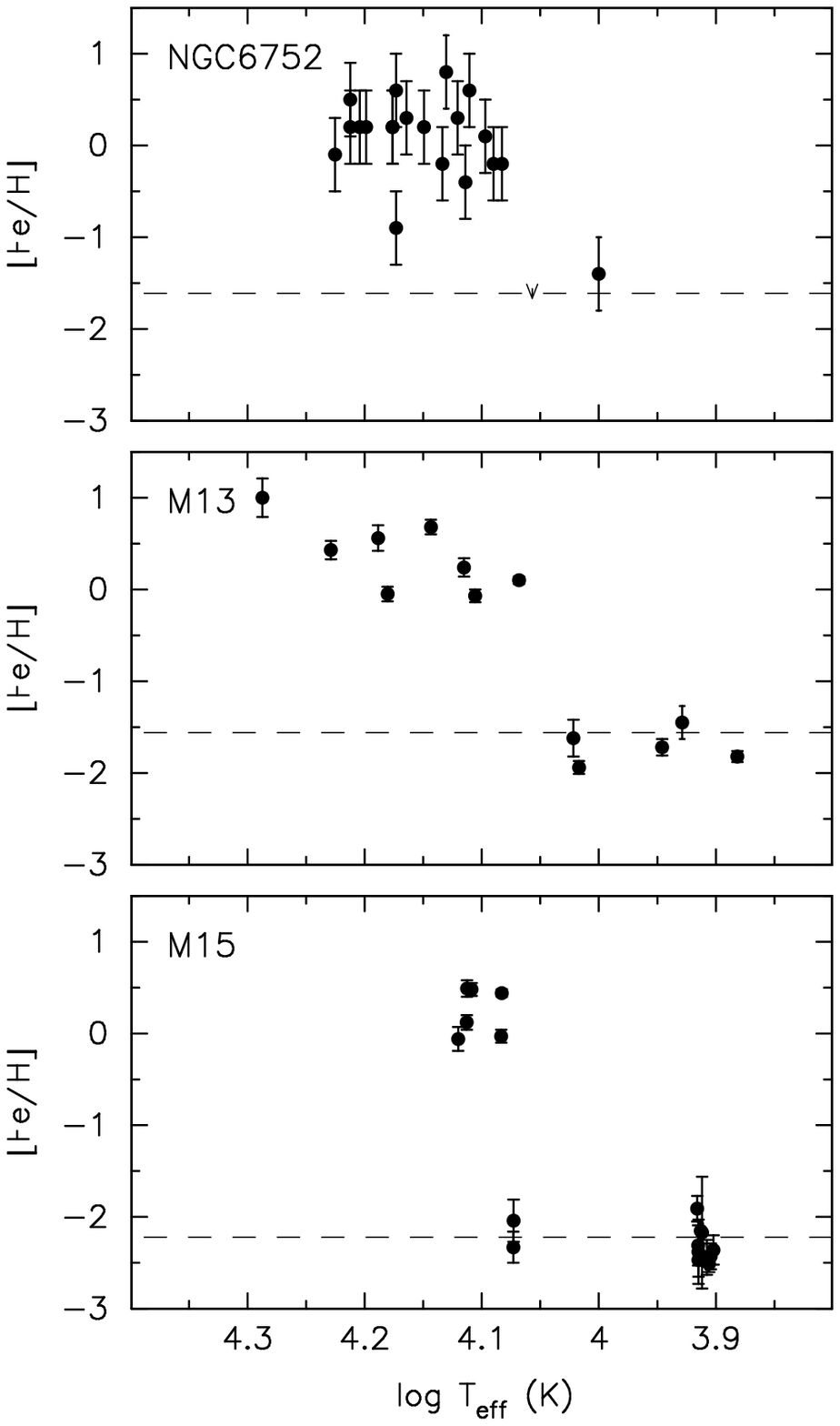}{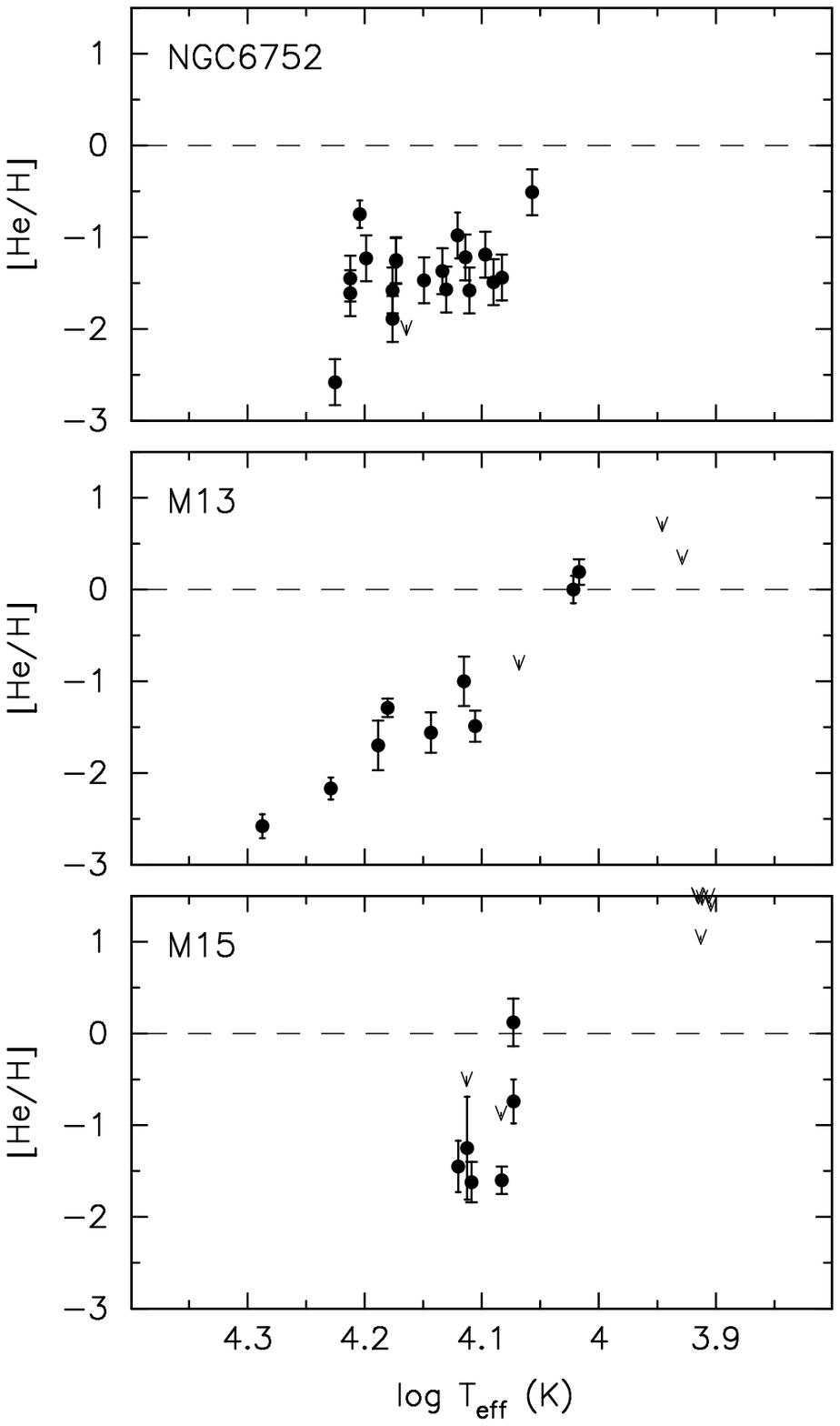}
\caption{Iron and helium abundances of BHB stars as a function of HB position. Horizontal dashed lines
indicate the expected abundance level for stars in each cluster, and inverted carets represent upper
bounds on abundance. For $T_{\rm eff} > 11500 \, {\rm K}$ ($\log T_{\rm eff} > 4.06$), iron is
strongly enhanced and helium is depleted. Data are from Moehler et al. (1999) and Behr et al. (1999,
2000b).}
\end{figure} 

Metallicity differences also appear among HB stars, as first discovered by Glaspey et al. (1989), who
analyzed two BHB stars in NGC~6752. Their cooler target, with $T_{\rm eff} \simeq 10000 \, {\rm
K}$, had ``normal'' iron and helium abundances for the cluster, but the hotter star, at $T_{\rm eff}
\simeq 16000 \, {\rm K}$, showed an iron abundance 50 times greater than expected, and a helium
abundance 40 times lower. More recent observations covering a larger range of $T_{\rm eff}$ in
NGC~6752 (Moehler et al. 1999), M13 (Behr et al. 1999), and M15 (Behr et al. 2000b) confirm these
abundance deviations, and reveal the presence of an abundance ``jump'' at $T_{\rm eff} \simeq 11500
\, {\rm K}$, closely coincident with the transition in rotation velocities. Redwards of the jump
temperature, stars show normal cluster abundances, but bluewards of the jump, the stars are strongly
enhanced in iron and most other metals, and depleted in helium, as illustrated in Figure 2. The
magnitude of the metal enhancement varies from element to element: the iron abundance increases by
1.5 to 2.0 dex (30 to 100 times), phosphorus increases by 3.0 to 3.5 dex, while magnesium is almost
entirely unchanged, remaining at metal-poor cluster abundance levels over the entire observed
temperature range.

\section{Diffusion mechanisms}

Atomic diffusion appears to be the best explanation for the large variations in chemical abundances
seen among the hotter BHB stars. To explain the anomalously low helium abundances found among
evolved halo stars, Greenstein et al. (1967) proposed that helium atoms sink under the influence of
gravity, diffusing from the photosphere into the stellar interior, so that the observed helium abundance
is lowered. This mechanism would be effective only if the atmosphere were very stable, since
convective or rotational mixing would completely erase the diffusion gradient. Michaud et al. (1983)
then pointed out that these same stable atmospheric conditions would also permit levitation of metal
atoms. Driven by radiation pressure, atomic species with sufficiently large radiative cross-sections
could be lifted from the stellar interior to the photosphere, such that large metal enhancements appear.
Recent models of similar diffusion mechanisms in chemically peculiar main-sequence Ap stars,
by Richer et al. (2000), show abundance patterns that are qualitatively similar to those we observe:
enhancement of iron, even stronger enhancement of phosphorus, no change in magnesium, and depletion
of helium. Hopefully, these sorts of models will soon be applied to the specific case of hot BHB stars, to
see whether we can quantitatively reproduce the spectroscopic observations, and thus verify that
radiative levitation and gravitational settling are responsible for the BHB abundance variations.

We must also explain why the amount of metal enhancement and helium depletion should change so
abruptly as a function of position along the HB. One likely possibility is that the disappearance of
surface convection at a threshold $T_{\rm eff}$ is the ``switch'' which regulates the effectiveness of
diffusion. BHB atmosphere/envelope models by Sweigart (2001) show that thin regions of convection
due to H I and He I lie at (or slightly below) the stellar surface for cooler BHB stars, but when
the model $T_{\rm eff}$ exceeds $12000 \, {\rm K}$, these convection zones disappear, and the
atmosphere is fully radiative in its upper layers. Thus, for stars hotter than the threshold
temperature, diffusion alters the photospheric abundance pattern, while the stars cooler than threshold
remain well-mixed, and no diffusion can take place. It has also been suggested that the abrupt change in
stellar rotation characteristics may determine whether diffusion is effective. If a star is rotating very
quickly, meridonal circulation currents can indeed provide enough mixing to prevent diffusion, but
since we see plenty of slowly-rotating cool BHB stars with normal abundances, rotation is probably not
a critical factor in this regime.

Strong metal enhancement and helium depletion of a star's photosphere will alter the atmospheric
opacities and emergent spectral energy distribution, changing the star's photometric properties. Even
before the existence and magnitude of the metallicity jump was fully known, both Caloi (1999) and
Grundahl et al. (1999) proposed that the photometric gaps and jumps that appear along the BHB in many
clusters might be due to the sudden onset of diffusion-driven metal enhancements, and initial
quantitative tests by Grundahl showed that the observed jumps in Stromgren $u$-band luminosities
could be partially reproduced in synthetic photometry by increasing the model atmosphere metallicities.
More comprehensive calculations by Hui-Bon-Hoa et al. (2000), which explicitly included element
diffusion and stratification, found that metal-enhanced photospheres are consistently bluer in
$U-V$ and Stromgren $u-y$ colors than normal metal-poor photospheres, with luminosity
jumps similar in size to those observed. The metallicity jump, then, can explain the Stromgren
$u$-jump and {\it some} of the photometric gaps, such as those around $11000 \, {\rm K}$ in M13
and M80, although cooler gaps like that in M15, and the hotter ones found in NGC~2808, require
alternative explanations. The anomalously low $\log g$ values measured for hot BHB stars also be
partly explained by the metal enhancements, although some degree of helium mixing may be necessary
as well (Moehler et al. 2000).

\section{Possible explanations of the rotation distribution}

The observed distributions of stellar rotation velocities are somewhat more difficult to explain than
the abundance variations. Several possible mechanisms have been suggested, but there is no compelling
theoretical or observational evidence to indicate which (if any) is correct. Any complete explanation of
BHB star rotation must explain three main aspects of the observations: (1) what causes the bimodal
distribution of rotation velocities among the cooler stars, (2) why does this bimodality appear only
among $\sim\onehalf$ of the clusters observed, and (3) why do the hotter, metal-enhanced stars all
rotate at very low $v_{\rm rot}$?

Peterson et al. (1983a) and Pinsonneault et al. (1991) suggested that the magnetic braking of a G-type
dwarf might only affect its envelope, while its core continues to spin rapidly, retaining much of the
star's primordial angular momentum. After the star arrives on the HB, the core couples to the
envelope, spinning up the surface layers to the observed velocities. Sills \& Pinsonneault (2000)
subsequently refined this scenario, pointing out that if core-envelope coupling takes an appreciable
fraction of the star's HB lifetime, then this could explain the observed bimodality in surface rotation
--- the fast-rotating stars are those that have been on the HB longer, and have had more time to spin
up. Futhermore, they claim, the gradients in mean molecular weight created by helium settling could
delay, or even prevent, the transfer of angular momentum from core to envelope, explaining the slow
rotation among the hotter stars. This hypothesis predicts that the fast-rotating stars, having spent
more time on the HB, will have evolved to slightly higher luminosities. Initial tests (Behr 2000c;
Recio-Blanco et al. 2002) show no such correlation between higher rotation and higer luminosity,
although the sample sizes and photometric quality are insufficient to rule out this possibility.
Helioseismology results (Corbard et al. 1997; Effdarwich et al. 2002) indicate that fast core rotation
is not present in the Sun, although some of the Sills \& Pinsonneault models demonstrate that faster
internal rotation can develop on the RGB as the core shrinks, even given solid-body rotation on the
main sequence, so some degree of internal rotation is still a possibility.

Mass loss on the HB provides another means for creating uniform slow rotation among the hot
metal-enhanced stars. Vink \& Cassisi (2002) describe models of BHB mass loss and resulting loss in
angular momentum, and find that mass loss rate is strongly dependent upon photospheric metallicity.
They propose that the large metal enhancements that appear for $T_{\rm eff} > 11500 \, {\rm K}$
result in a large increase in mass loss, which rapidly removes angular momentum from the star, such
that the hotter BHB stars are all slow rotators.

The ``excess'' angular momentum in the fast-rotating BHB stars could come from an external source,
instead of originating within each individual star. Although stellar collisions are quite rare, even in
dense cluster centers, close tidal encounters could ``spin up'' a star, assuming the impact parameter
was in the right range. Denser cluster environments do appear to affect the evolution and properties of
individual stars, as discussed by Fusi Pecci et al. (1993), Buonanno et al. (1997), and Testa et al.
(2001), who note a correlation between higher cluster core densities and the presence of long blue
tails in clusters' CMDs. According to this model, stars in denser environments are more prone to tidal
encounters, which enhance mass loss (either via direct tidal stripping, or spin-up), resulting in smaller
hydrogen envelopes on the HB, and thus bluer stars.

Alternatively, a stellar companion might provide the additional angular momentum, either through tidal
synchronization of the orbital and rotation periods, or via a merger. Stellar mergers are known to
happen in GCs, forming fast-rotating blue straggler stars, but the subsequent evolution of these stars
would probably {\it not} put them on the same HB locus as ordinary single stars. No extended radial
velocity monitoring of BHB stars has been performed, so we cannot test for binarity via ``Doppler
wobble,'' but the velocity dispersions of BHB stars closely match those of each cluster population at
large (Peterson et al. 1983b), suggesting that large radial velocity variations are not present.
Furthermore, BHB stars do not appear to be significantly concentrated towards cluster centers, as
many binary star types are. Therefore, stellar binarity does not appear to be common enough to
explain the fast rotators.

Smaller companions might still be plausible, however. If a large planet in a close orbit were absorbed
by an expanding RGB envelope, it could easily provide sufficient angular momentum to create a
fast-rotating BHB star. This scenario was initially suggested by Peterson et al. (1983a), and recent
calculations by Soker \& Harpaz (2000) and Livio \& Soker (2002) show that it is quantitatively
plausible. Some ``51 Peg-type planets,'' with $a \simeq 0.05 \, {\rm AU}$, have been found orbiting
stars in the solar neighborhood, but they appear to be quite uncommon in globular clusters, according
to the 47 Tucanae transit survey of Gilliland et al. (2000). Planets in wider orbits ($a = 0.2$--$0.5 \,
{\rm AU}$) cannot be ruled out, and would still be absorbed by the red giant, so the planet hypothesis
remains tenable, but there currently exists no independent evidence that planets are sufficiently
common in GCs to create the observed fast-rotating BHB population.

\section{Summary}

Several independent spectroscopic investigations show that photospheric abundances and stellar
rotation rates of BHB stars vary significantly as a function of temperature. All BHB stars hotter than
$T_{\rm eff} \simeq 11500 \, {\rm K}$ show metal enhancements of 1.5 to 3.5 dex, and helium
depletions of $\sim 2$~dex. These abundance ``anomalies'' are most likely due to atomic diffusion
processes --- radiative levitation of the metals, and gravitational settling of helium --- in the stable
non-convective atmospheres of the hotter stars. The sudden onset of metal enhancements, possibly due
to the disappearance of surface convection, can alter the spectrophotometric properties of the star,
thus explaining some, but not all, of the photometric gaps and jumps which have been found in cluster
CMDs.

The hotter, metal-enhanced BHB stars all appear to rotate slowly, $v \sin i < 8 \, {\rm km} \, {\rm
s}^{-1}$, in contrast to the cooler ($T_{\rm eff} \simeq 8000$--$11500 \, {\rm K}$) BHB stars,
which show a range of rotation velocities. In some clusters (and the field population),
approximately one-third of the cooler stars are rotating at $v_{\rm rot} \simeq 35$--$40 \, {\rm
km} \, {\rm s}^{-1}$, considerably faster than one would expect for the progeny of slow-rotating
main-sequence G stars, while the other two-thirds have more modest rotation rates, $v_{\rm rot}
\simeq 15$--$20 \, {\rm km} \, {\rm s}^{-1}$. In other clusters, only the slower-rotating stars
appear. This distribution of rotation rates might be due to the evolution of internal angular momentum
within each star, or may instead be the result of dynamical interactions with other stars or substellar
objects within the cluster.

Further BHB rotation observations spanning a range in metallicity, cluster concentration, and cluster
size will help to explain the observed $v \sin i$ distributions. In particular, we need to determine why
the presence and peak $v_{\rm rot}$ of a fast-rotating population differs from cluster to cluster;
if we can correlate stellar rotation with characteristics of the parent clusters, this would provide
additional insights into the origin of the fast rotation. With large samples in individual clusters, we
could also look for differences in kinematics and radial distribution between the slow and fast rotators.
More extensive studies of the field BHB population are also necessary, as these stars exist in a very
different dynamical environment from their cluster analogs. The issue of binarity should be tested
more directly, with long-term monitoring for radial velocity variations, especially in light of the large
binary fractions found among hot evolved halo stars (Saffer et al. 2000; Maxted et al. 2001) and
metal-poor field RGB stars (Carney et al., this volume), and future searches for planets around
metal-poor GC stars will let us determine whether swallowing of planetary companions is a plausible
source of angular momentum. We look forward to continued refinement of stellar models which track
the change in internal angular momentum profile as a result of evolution, binary and tidal interaction,
and merger events.

\end{document}